\documentclass[12pt,osajnl2,preprint,showpacs,amsfonts,amssymb,amsmath,eqsecnum,nofootinbib,floatfix,superscriptaddress]{revtex4}  

\usepackage{graphicx}
\usepackage{epsf}                
\usepackage{psfrag}
\usepackage{bm}

\usepackage{dcolumn}

\usepackage{etoolbox}
\apptocmd{\sloppy}{\hbadness 10000\relax}{}{}

\newcommand{\myvec}[1]{\vec{#1}\,}

\newcommand{\beq}{\begin{equation}}
\newcommand{\eeq}{\end{equation}}
\newcommand{\bea}{\begin{eqnarray}}
\newcommand{\eea}{\end{eqnarray}}

\begin{document}

\preprint{LIGO-P050050-00-R}

\title{A duality relation between non-spherical mirror optical cavities and its application to gravitational-wave detectors}

\author{Juri Agresti}
 \affiliation{Istituto di Fisica Applicata ``Nello Carrara'' - CNR, 50019 Sesto Fiorentino (FI), Italy}
\affiliation{LIGO Laboratory, California Institute of Technology,
Pasadena, CA 91125}
\author{Yanbei Chen}
\affiliation{Theoretical Astrophysics, California Institute of Technology, Pasadena, CA 91125}
\affiliation{Max-Planck-Institut f\"ur Gravitationsphysik (Albert-Einstein-Institut), D-14476 Golm, Germany}
\author{Erika D'Ambrosio}
\affiliation{LIGO Laboratory, California Institute of Technology, Pasadena, CA 91125}
\author{Pavlin Savov}
\affiliation{Theoretical Astrophysics, California Institute of Technology, Pasadena, CA 91125}

\date{\today }

\begin{abstract}
 In this paper, we analytically prove a unique
duality relation between the eigenspectra of paraxial optical
cavities with non-spherical mirrors: a one-to-one mapping between
eigenmodes and eigenvalues of cavities deviating from flat mirrors
by $h(\myvec{r})$ and cavities deviating from concentric mirrors
by $-h(\myvec{r})$, where $h$ need not be a small perturbation. We
then illustrate its application to optical cavities, proposed for
advanced interferometric gravitational-wave detectors, where the
mirrors are designed to support beams with rather flat intensity
profiles over the mirror surfaces.  This unique mapping might be
very useful in future studies of alternative optical designs for
advanced gravitational waves interferometers or experiments
employing optical cavities with non-standard mirrors.
\end{abstract}

\ocis{070.2580,080.4228,260.0260,230.0230,140.0140}

 \maketitle

\section{\label{sec:intro} Introduction}

Laser Interferometer Gravitational-wave Observatory
(LIGO)~\cite{LIGO} and other long baseline detectors, are formed
by high-Finesse Fabry-Perot arms in order to increase the
circulating optical power and to enhance sensitivities by
suppressing shot noise. LIGO interferometers, as well as the
baseline design for Advanced-LIGO detectors ~\cite{LIGO2}, all use
spherical mirrors and fundamental Gaussian mode. Mirrors thermal
noise is expected to be the dominant source of noise in the most
sensitive frequency band of second-generation, ground-based
gravitational-wave detectors. Different shapes of beam have been
proposed for reducing this noise, such as rather flat {\it
mesa}-like~\cite{Willems} mode called Mesa
beams~\cite{OT1,mbi,osv,EA,Tarallo}, conical modes~\cite{BKC} and
high order Laguerre–Gauss modes~\cite{LG1,LG2}. In particular, the
former two approaches and more general optimized beam
profiles~\cite{Pierro}, require the use of non-spherical mirrors
in the Fabry-Perot optical cavities. The research on Mesa beams
and non-spherical mirrors supporting them was very active in the
past few years and led to the discovery of the duality relation as
described in the following.

O'Shaughnessy, Thorne and Agresti~\cite{osv,mbi,tnA}, calculated
that the thermal fluctuations of mirror surfaces are better
averaged over by Mesa beams with respect to Gaussian ones. The
corresponding optical design has shown a strong tilt
instability~\cite{SandV} and Thorne has proposed a different
version of the Mesa beam, that is supported by nearly {\it
concentric} and opportunely shaped mirrors; this new version
provides the same intensity profile at the cavity mirrors (and
thus the same thermal noise), but imply a weaker tilt instability
(even weaker than cavities with nearly concentric spherical
mirrors analyzed by Sigg and Sidles~\cite{sigg,sidles})
--- as calculated by Savov and Vyatchanin~\cite{SandV}. A general
method to design a family of optical cavities from nearly flat  to
nearly concentric ones, has been devised by Bondarescu and
Thorne~\cite{BT} and the resulting fundamental mode, called
hyperboloidal beam, was later studied in more details as an
alternative to Mesa beams~\cite{Galdi1,Lundgren}.

Mesa beams are constructed by coherently overlapping Gaussian
beams, with either(i) translated parallel axes, or (ii) axes in
different directions but sharing a common midpoint~\cite{BT}.
Mirror shapes which support such beams as fundamental modes are
derived from the phase fronts at the mirror locations, with case
(i) corresponding to Mexican-hat mirrors, and case (ii)
corresponding to the nearly-concentric version. Using the
resulting optics profile, higher-order optical modes and
eigenfrequencies of the designed cavities must be calculated by
solving an eigenvalue problem, which has been done for nearly-flat
cavities by O'Shaughnessy and Thorne~\cite{osv,mbi}, and for
nearly-concentric cavities by Savov and Vyatchanin~\cite{SandV}.
During his numerical work, Savov discovered that the deviation of
nearly-concentric Mexican-hat mirrors from concentric surfaces is
exactly the opposite of the deviation of nearly-flat Mexican-hat
mirrors from flat surfaces; he also found that the corresponding
higher modes of these cavities all have the same intensity
profiles, and that there is a one-to-one mapping between their
eigenvalues. Following this numerical analysis Bondarescu
conjectured a general {\it duality relation} between axisymmetric
cavities with two identical mirrors facing each other:  cavities
with mirrors deviating by $-h(|\vec r\,|)$ from concentric
surfaces ({\it nearly concentric mirrors}) will support modes with
the same intensity profiles and related eigenvalues as cavities
with mirrors deviating by $h(|\vec r\,|)$ from flat surfaces ({\it
nearly flat mirrors}). It should be noted that the deviation
$h(\vec r\,)$ is {\it not} required to be infinitesimal, it can
change the mirror shape arbitrarily {\it as long as the paraxial
approximation is still satisfied}. Here and henceforth in the
paper a 2-D vector $\vec r$  has been used to indicate each point
on planes orthogonal to the the cavity axis. While such a duality
relation is well-known between cavities with spherical mirrors,
i.e., those with $h(\vec r)\sim \alpha \vec r\,^2$~(for example
see~\cite{GorKog,ER,siegman2,FL,KL}), to our best knowledge no
such relations had been established between generic cavities.

In this paper, we prove this remarkable correspondence
analytically, for a even broader category of cavities: those whose
mirror shapes remain invariant under the parity operation,
identified as spatial reflection in the two dimensional $\myvec
r$-space (which is also equivalent to a $180^{\circ}$ rotation
around the cavity axis). Eigenmodes of such cavities can be put
into eigenstates of parity, and we show that all corresponding
eigenmodes of dual cavities have the same intensity profiles at
the mirrors, with their eigenvalues satisfying
\begin{equation}
\label{eq:intro:dual}
 \gamma_{\rm c}^k=(-1)^{p_k+1} e^{-2ikL} (\gamma_{\rm f}^k)^*\,,
\end{equation}
where $(-1)^{p_k}$ is the parity of the $k$th eigenmode;
subscripts c and f denote nearly concentric and nearly flat
mirrors, respectively.

We will give two alternative proofs of this duality relation. The
first one relies on the geometrical properties of the propagator
from mirror to mirror. In this description the eigenfunctions are
field amplitudes at mirror surfaces, and we see right away that
the corresponding eigenstates have the same intensity profiles
there. The second proof is based on the ``center-to-center''
propagator. The center-of-the-cavity fields are the eigenstates
and the correspondence relation is manifested by a two-dimensional
Fourier transform, that univocally relates the dual cavities.

This paper is organized as follows. In Sec.~\ref{sec:proof1} we
report the first proof; in Sec.~\ref{subsec:proof1:cartesian}, the
Cartesian coordinates are used and some general features of the
eigenproblem are described; in
Sec.~\ref{subsec:proof1:cylindrical}, the cylindrical coordinates
are used, and the case of axisymmetric resonators is studied.
Section~\ref{sec:proof2} contains the second proof and the 2-D
Fourier transform relation between the center-of-the-cavity
eigenmodes of dual cavities. Section~\ref{sec:examples}
specializes to the case of Mexican-hat cavities. When the
nearly-flat and the nearly-concentric mirrors are implemented in
the system, the corresponding Mesa beams are connected by Fourier
transform, as we report in Sec.~\ref{subsec:examples:mapping}. In
Sec.~\ref{subsec:examples:profile}, plots and analytical forms are
provided, for the amplitude distributions at the center of the
cavity and at the mirror surfaces; in
Sec.~\ref{subsec:examples:tilt}, we address the tilt instability
of the nearly concentric Mexican-hat resonator and show how easily
it can be analyzed, applying the duality relation to the results
obtained for the nearly flat Mexican-hat cavities~\cite{osv,mbi}.
We comment and review the implications of the general duality in
Sec.~\ref{sec:conclusion}.

\section{\label{sec:proof1} Analytical proof for mirror-to-mirror propagation}

\subsection{\label{subsec:proof1:cartesian} In the Cartesian coordinate system}

\begin{figure}[htbp]
\begin{center}
\includegraphics[width=0.4\textwidth]{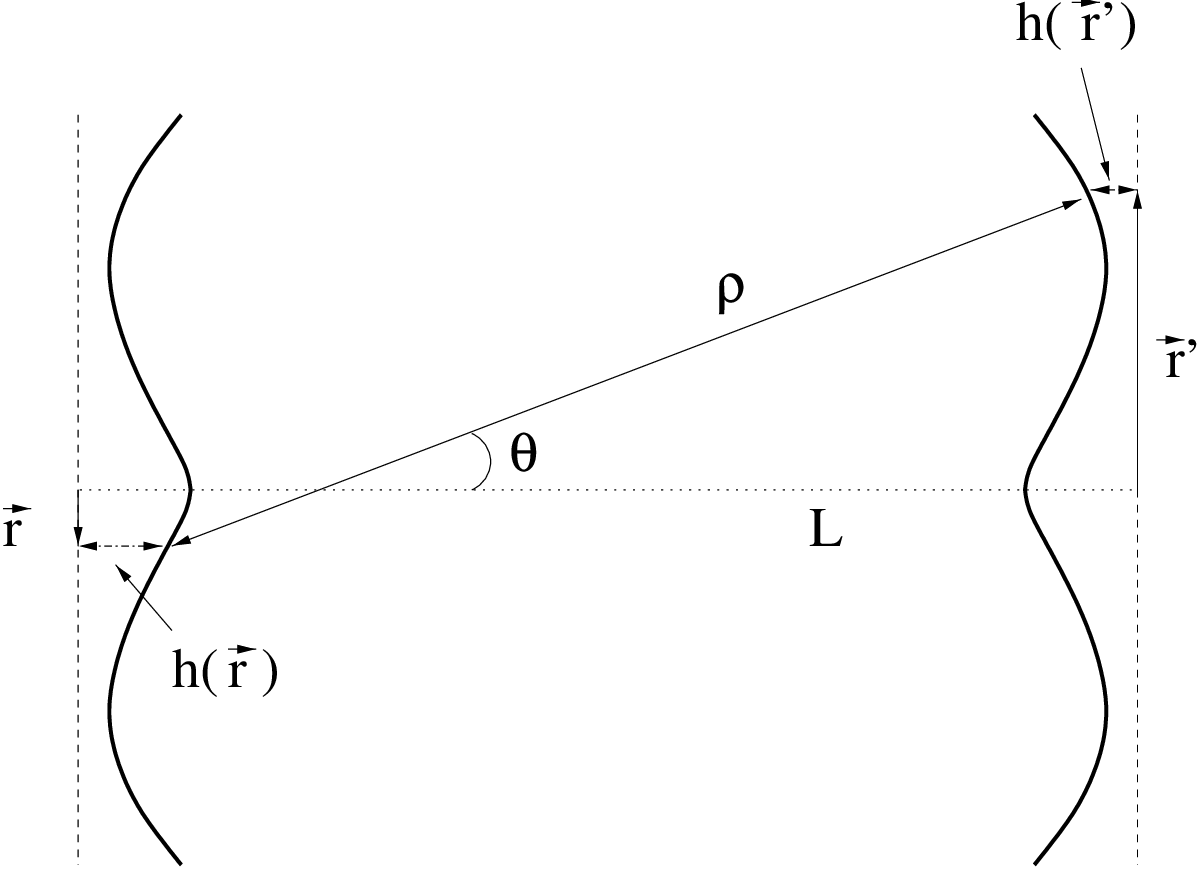}
\end{center}
\caption{Symmetric Nearly Flat Mirrors.}
\label{fig:flat}
\end{figure}

In this section we focus on field distributions on mirror
surfaces, and restrict ourselves to cavities with two identical
mirrors facing each other. The extension to asymmetric cavities is
presented in Appendix~A. We adopt the Fresnel-Kirchoff diffraction
formula to propagate fields from mirror surface to mirror surface
(see e.g.~\cite{FL}).  In this formalism, the field amplitude
$v_1(\myvec r')$ on the surface of mirror 1 propagates into
\begin{equation}
v_2(\vec r)=\int\mbox{d}^2\vec r\,'\;{\cal K}(\vec r,\vec r\,')\,v_1(\vec r\,')
\end{equation}
on mirror 2, via the propagator \beq {\cal K}(\vec r,\myvec r')
=\frac{i k}{4 \pi \rho}(1 + \cos\theta)e^{-i k \rho} \qquad\qquad
k=\frac{2\pi}{\lambda}\,, \eeq from $\myvec r'$ (on mirror 1) to
$\myvec r$ (on mirror 2), where $\rho$ denotes the (3-D) spatial
distance between these two points and $\theta$ stands for the
angle between the cavity axis and the reference straight line, as
is illustrated in Fig.~\ref{fig:flat}. We know that the
Fresnel-Kirchoff integral eigenequation\begin{equation}
\label{eq:eigen} \gamma \, v(\vec r)=\int\mbox{d}^2\vec
r\,'\;{\cal K}(\vec r,\vec r\,')\,v(\vec r\,')
\end{equation}
univocally determines the eigenmodes $v$ and eigenvalues $\gamma$ of the cavity.

Applying the paraxial approximation
\begin{equation}
 \label{eq:rho}
\theta \approx 0\,,\quad \rho \approx L +\frac{|\myvec r -\myvec r'|^2}{2L}-h(\myvec r) - h(\myvec r')\,,
\end{equation}
and we can use
\begin{equation}
{\cal K}^{h}_{\rm f}(\vec r,\vec r\,')=\frac{ik}{2\pi L}e^{-ikL}
e^{ikh(\vec r)}e^{-\frac{ik}{2L}|\vec r-\vec r\,'|^2}e^{ikh(\vec r\,')}\,.
\label{eq:aa}
\end{equation}
in the integral eigenequation.

\begin{figure}[ht]
\begin{center}
\includegraphics[width=0.4\textwidth]{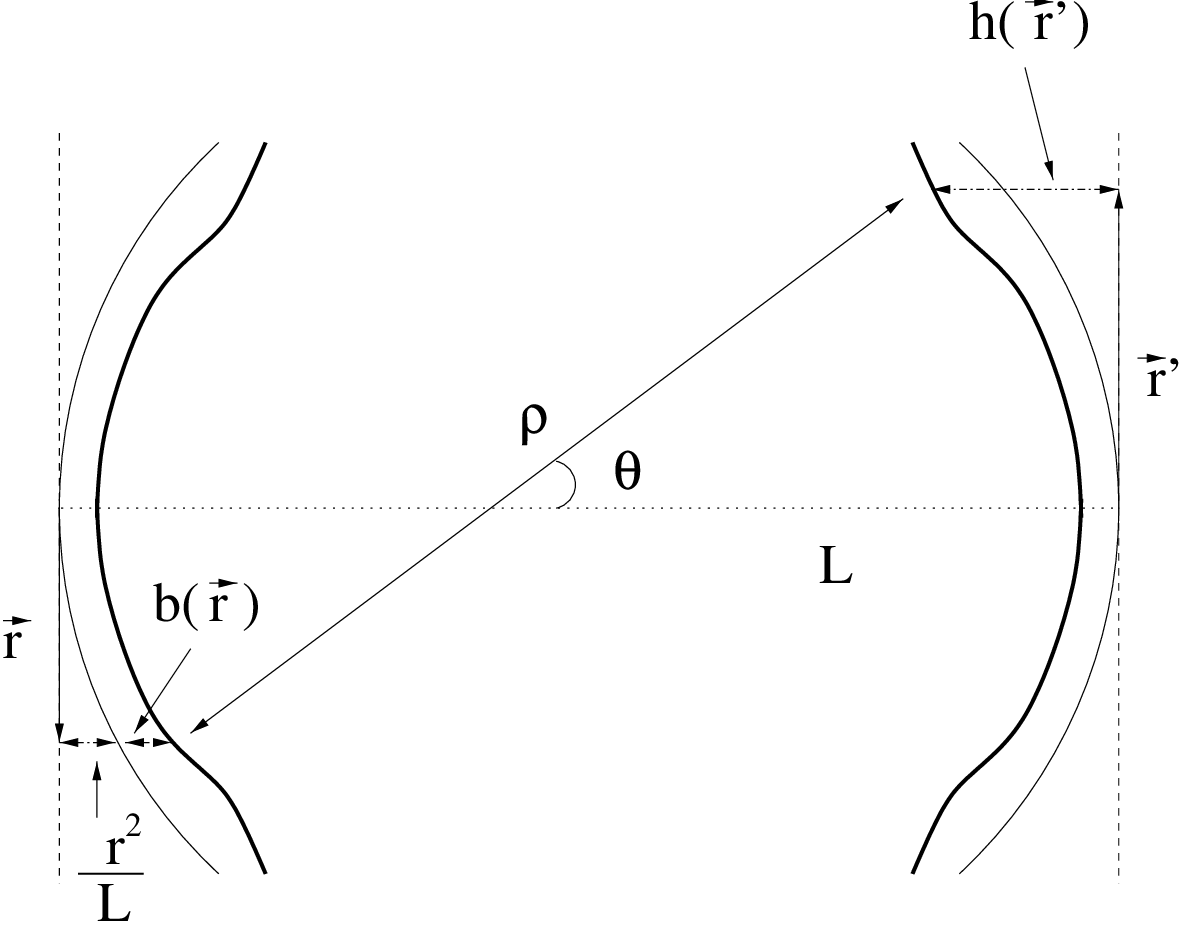}
\end{center}
\caption{Symmetric Nearly Concentric Mirrors.} \label{fig:sfer}
\end{figure}

Here the mirror surfaces deviate by $h(\myvec r)$ from a flat
reference, and the subscript {\it f} is used to reflect this
convention. From here on, we will also refer to $\mathcal{K}_{\rm
f}^h$ as the {\it nearly flat propagator}. We now consider two
slightly deformed concentric mirrors (see Fig.~\ref{fig:sfer}) so
that the mirrors height with respect to the flat reference surface
is
\begin{equation}
h(\myvec{r})=\vec r\,^2/L\,+b(\myvec{r})\, , \label{eq:conv}
\end{equation}
where the height $b(\myvec{r})$ is the deviation from the concentric spherical
surface (note that concentric spherical mirrors have their radii of curvature equal to $L/2$, and thus surface height $r^2/L$).
Inserting Eq.~\eqref{eq:conv} into Eq.~\eqref{eq:rho}, we obtain the propagator for a {\it nearly-concentric} cavity,
\begin{equation}
{\cal K}^{b}_{\rm c}(\vec r,\vec r\,')=\frac{ik}{2\pi L}e^{-ikL}
e^{ikb(\vec r)} e^{+\frac{ik}{2L}|\vec r+\vec r\,'|^2}e^{ikb(\vec
r\,')}\, \label{eq:aaconc}
\end{equation}
 We use the term {\it nearly concentric} propagator for ${\cal K}^{b}_{\rm c}(\vec r,\vec r\,')$.
Although we use the terms {\it nearly-flat} and {\it
nearly-concentric}, $h$ and $b$ are not required to be small; in
fact, they can represent any deviation from perfectly flat and
concentric spherical mirrors.

Now let us consider mirrors that are then invariant under parity, i.e., those in which we also have
\beq
h(\vec r)= h(-\vec r)\,,\qquad b(\vec r)=b(-\vec r)\,.
\eeq
so that $\mathcal{K}_{\rm f,\,c}$ are both invariant under a spatial reflection
\begin{equation}
\left\{\vec r, \vec r\,' \right\}\leftrightarrow\left\{-\vec r,-\vec r\,' \right\}
\end{equation}
and therefore, we have
\begin{equation}
\label{commute}
\mathcal{P}\mathcal{K}=\mathcal{K}\mathcal{P}\,,
\end{equation}
where we have defined
\begin{equation}
\mathcal{P}v(\vec{r})=v(-\vec{r})\,.
\end{equation}
for two dimensional reflection. Equation~\eqref{commute} implies
that all eigenmodes can be put into forms with definite parity. We
derive the following relation between nearly flat and nearly
concentric propagators, as constructed:
\begin{equation}
\label{eq:correspondence}
\left[\mathcal{K}_{\rm f}^h(-\vec r,\vec r\,')\right]^*
=-e^{2ikL}\mathcal{K}_{\rm c}^{-h}(\vec r,\vec r\,')\,,
\end{equation}
that is equivalent to:
\begin{equation}
\label{eq:duality}
\mathcal{P} \left[\mathcal{K}_{\rm f}^{h}\right]^*
=-e^{2ikL} \mathcal{K}_{\rm c}^{-h}\,.
\end{equation}
Suppose we have an eigenstate $v_{\rm f}$ of $\mathcal{K}_{\rm
f}^{h}$, i.e., an eigenstate of a cavity with mirror deviating by
$(+h)$ from flat surface, and we compute its eigenvalue
$\gamma_{\rm f}$ and know the parity eigenvalue $(-1)^p$:
\begin{eqnarray}
\label{eq:v1}
\mathcal{K}_{\rm f}^{h}v_{\rm f} &=&\gamma_{\rm f}\, v_{\rm f}\,,\\
\label{eq:v2}
\mathcal{P}v_{\rm f}&=&(-1)^p v_{\rm f}\,.
\end{eqnarray}
By applying Eqs.~\eqref{eq:duality}--\eqref{eq:v2}, we derive the correspondance\begin{equation}
\mathcal{K}_{\rm c}^{-h} v_{\rm f}^*=e^{-2ikL}(-1)^{p+1}\gamma_{\rm f}^*v_{\rm f}^*\,.
\end{equation}
which identifies $v_{\rm c} \equiv v_{\rm f}^*$ as the
corresponding eigenstate of $\mathcal{K}_{\rm c}^{-h}$, that is
eigenstate of the corresponding resonator we denote the {\it
dual}. The eigenvalue is $\gamma_{\rm c}\equiv
e^{-2ikL}(-1)^{p+1}\gamma_{\rm f}^*$. We also induce that the
parity is still $(-1)^p$. The reverse is straightforward and the
result is an established one-to-one correspondence between dual
cavities. We summarize this mapping in Table~\ref{tab:duality:EJ}.
It is obvious to note that that the corresponding eigenstates,
$v_{\rm f}$ and $v^*_{\rm f}$,  have the same intensity profiles
on the mirror surfaces; for infinite mirrors, we know $v_{\rm
f}(\vec r)$ is real-valued (see Appendix~\ref{app:b}), so it is an
eigenstate of the dual configuration itself.

\begin{table}[t]
\centerline{
\begin{tabular}{r|c|c}
\hline\hline
& Nearly Flat & Nearly Concentric\\
\hline
Kernel & $\mathcal{K}_{\rm f}^{h}$ & $\mathcal{K}_{\rm c}^{-h}$\\
Eigenstate & $v_{\rm f}$ & $v^*_{\rm f}$ \\
Parity & $(-1)^p$ & $(-1)^p$ \\
Half-trip eigenvalue & $\gamma_{\rm f}$ & $e^{-2ikL}(-1)^{p+1}\gamma_{\rm f}^*$ \\
\hline
Round-trip eigenvalue &  $\eta_{\rm f}$ & $e^{-4ikL}\eta_{\rm f}^*$
\\
\hline\hline
\end{tabular}
} \caption{\label{tab:duality:EJ} Correspondence of propagation
kernels, eigenstates, parities, and eigenvalues between dual
configurations.}
 \end{table}

 For cavities with identical mirrors facing each other, the full, round-trip propagator is just the square of the half-trip one. From Eqs.~\eqref{eq:correspondence} and \eqref{commute}, we have
\begin{equation}
\left[\left[\mathcal{K}_{\rm f}^h\right]^2\right]^*=e^{4ikL}\left[\mathcal{K}_{\rm c}^{-h}\right]^2
\end{equation}
which means that the same duality correspondence exists between eigenstates of the full propagator, with their eigenvalues related by
\begin{equation}
\eta_{\rm c} = e^{-4ikL}\eta_{\rm f}^*\,.
\end{equation}

Note that when $h(\vec r)=r^2/(2L)$ the two dual cavities are identical to each other.
Using the relation that links the eigenvalues of two dual resonators, we
can determine the spectrum
$$\gamma_c=\pm e^{-2ikL}\gamma_{f}^{*}=\gamma_f=e^{-ikL+in\pi/2}$$
where $n\in{\cal N}$. The resulting separation between the eigenvalues is the Gouy phase
$$e^{i\theta_G}=e^{i\arccos(1-L/R)}\qquad R=L$$
computed for confocal resonators \cite{siegman2,FL,KL}.

\subsection{\label{subsec:proof1:cylindrical} Specializing to cylindrical mirrors}

In most practical applications cavity mirrors have cylindrical
shapes: $h(\myvec{r})=h(|\myvec{r}|)$. This allows us to decouple
radial and azimuthal degrees of freedom, and simplify the
eigenvalue problem. We shall follow roughly the notation
of~\cite{KL}.

We adopt the cylindrical coordinate system:
\begin{equation}
\myvec{r} = r(\cos\varphi,\sin\varphi)\,.
\end{equation}
Since  $\mathcal{K}$ is now invariant under rotation along the $z$-axis, all eigenmodes can be put into eigenstates of rotation:
\begin{equation}
v(r,\varphi) = R(r) e^{-i m\varphi}\,,\quad m={\rm integer}\,.
\end{equation}
Inserting this into the eigenequation~\eqref{eq:eigen} and performing analytically the angular integration we obtain the radial eigenequation
\begin{equation}
\gamma_{nm} R_{nm}(r) = \int_0^a K_{\mathrm{f}(m)}^h(r,r')R_{nm}(r')r'dr'\,,
\end{equation}
where for each angular mode number $m$ we have indexed the radial eigenstates by $n$, and
\begin{equation}
\label{eq:radialK:f}
K_{\mathrm{f}(m)}^h(r,r')=\frac{i^{m+1}k}{L}J_m\left(\frac{kr r'}{L}\right)e^{ik\left[-L+h(r)+h(r')-\frac{r^2+r'^2}{2L}\right]}\,
\end{equation}
is a symmetric radial propagator, in the {\it nearly-flat}
description. Here we have used $\displaystyle J_n(z)= 1/(2\pi
i^n)\int_0^{2\pi} e^{iz\cos\varphi}e^{in\varphi}d\varphi$, where
$J_n(z)$ is the $n$th order Bessel function of the first kind.
Since $K_{\mathrm{f}(m)}^h(r,r')$ is symmetric, we obtain
orthogonality relations between radial eigenfunctions:
\begin{equation}
\int_0^a  R_{n_1m}(r) R_{n_2m}(r) rdr =\delta_{n_1n_2}\,.
\end{equation}

Using Eq.~\eqref{eq:conv} again, for a configuration with $b(r)$
correction from concentric spherical mirrors, we obtain the radial
kernel of the {\it nearly-concentric} description:
\begin{equation}
\label{eq:radialK:c}
K_{\mathrm{c}(m)}^b(r,r')=\frac{i^{m+1}k}{L}J_m\left(\frac{kr r'}{L}\right)e^{ik\left[-L+b(r)+b(r')+\frac{r^2+r'^2}{2L}\right]}\,.
\end{equation}
Comparing Eqs.~\eqref{eq:radialK:c} and \eqref{eq:radialK:f}, we obtain:
\begin{equation}
(-1)^{m+1}\left[K_{\mathrm{f}(m)}^h \right]^*  = e^{2ikL}K_{\mathrm{c}(m)}^{-h}\,.
\end{equation}
This is a radial version of Eq.~\eqref{eq:duality};  here we know explicitly that all $m$-eigenstates have parity $(-1)^m$.

Following a similar reasoning as done in the previous section,
{\it for each} angular mode number $m$, we can establish a
one-to-one correspondence between radial eigenstates of a
nearly-flat configuration to those of the dual configuration:
\begin{equation}
\left[R_{nm}\right]_{\rm c}=\left[R_{nm}\right]_{\rm f}^*\,.
\end{equation}
The mapping of the eigenvalues are given by
\begin{equation}
\left[\gamma_{nm}\right]_{\rm c} =(-1)^{m+1}e^{-2ikL}\left[\gamma_{nm}\right]_{\rm f}^*\,.
\end{equation}
Similarly, the round-trip eigenstates have the same correspondence, their eigenvalues related by
\begin{equation}
\left[\eta_{nm}\right]_{\rm c} =e^{-4ikL}\left[\eta_{nm}\right]_{\rm f}^*\,.
\end{equation}

\section{\label{sec:proof2} Analytical proof based on center-to-center propagation}

\subsection{\label{sec:main} Propagators for vacuum and mirror surfaces }

In this section, we focus on complex amplitudes of the optical
field on planes perpendicular to the optical axis (the $z$ axis).
An optical mode propagating along one direction of the optical
axis can be specified completely by the distribution of the field
on the $z={\rm const}$ plane. For example, we denote the optical
field on the plane $z=z_1$ by $v(\myvec{r},z_1)$, where
$\myvec{r}$ is the 2-D coordinate of the point on this plane.  The
effect of any linear paraxial optical system (including open
space, thin lenses and mirrors) with input plane $z_1$ and output
plane $z_2$ can be characterized by its {\it transfer operator},
$\mathcal{U}$, which takes the form of an integration kernel:
\begin{equation}
v(\myvec{r},z_2)=\int{d^{2}{\myvec{r}}'\,\mathcal{U}(\myvec{r},z_2;\myvec{r}',z_1)v(\myvec{r}',z_1)}\,.
\end{equation}
In particular, the operator that describes the paraxial propagation down a length $L$ in vacuum
is
\begin{equation}
\label{eq:GL}
\mathcal{G}_L(\myvec{r},\myvec{r}')=i\frac{k}{2\pi L} e^{-ikL}\exp\left[-ik\frac{(\myvec{r}-\myvec{r}')^2}
{2L}\right]\,.
\end{equation}
For a mode propagating in the $\pm z$ direction with field
(complex) amplitude distribution $v(\myvec{r}',z_1)$ at $z=z_1$,
the amplitude distribution on a surface described by height
$z(\myvec{r})=z_1 \mp h(\myvec{r})$ is given by
\begin{equation}
\label{eq:mirror}
v[\myvec{r},z(\myvec{r})] = e^{\pm ikh(\myvec{r})}v[\myvec{r},z_1]\,.
\end{equation}
Here we emphasize that the spatial point of interest is located
outside the $z=z_1$ plane, and that the 2-D vector  $\myvec{r}$
describes the projection of that point onto the $z=z_1$ plane.

From Eq.~\eqref{eq:mirror}, one deduces that the operator for
reflection off a perfect infinite mirror with shape $h(\vec{r})$
is
\begin{equation}
\label{Ropt}
\mathcal{R}_{[h(\myvec{r})]}(\myvec{r},\myvec{r}') \equiv -\delta(\myvec{r}-\myvec{r}') e^{2 i k h(\myvec{r})}.
\end{equation}

The minus sign in Eq.~\eqref{Ropt} is used because we use a
convention in which a phase shift by $\pi$ is gained upon
reflection. It is easy to verify that both $\mathcal{G}_L$ and
$\mathcal{R}_{[h(\myvec{r})]}$ are unitary operators.

\subsection{Analytical proof based on center-to-center propagation}

In this section we present an alternative  proof motivated from
the construction of the Mesa beams~\cite{osv,SandV}: (i) the
nearly flat configuration has its fundamental mode generated by
{\it spatial translation} of minimal Gaussian beams, while (ii)
the nearly concentric configuration is generated by {\it rotation}
(of propagation direction at the center of cavity) of minimal
Gaussian beams, or a translation in the momentum
$\myvec{k}$-space.  This had led us to speculate that the two sets
of eigenstates correspond to each other via  {\it Fourier
transform} (similar to the relation between position and momentum
space in quantum mechanics).

We will use the operator $\mathcal{G}_{L/2}$ [see
Eq.~\eqref{eq:GL}] which propagates the field forward by {\it
half} the cavity length. For simplicity we denote it by
$\mathcal{G}$:
\begin{eqnarray}
\label{defG}
\mathcal{G}(\myvec{r},\myvec{r}')&\equiv&i \frac{k}{\pi L} e^{- i k L/2} e^{ - i k \frac{(\myvec{r}-\myvec{r}')^2}{L}}\,.
\end{eqnarray}
Using $\mathcal{G}$ and $\mathcal{R}_{h(\myvec{r})}$ [defined in
Eq.~\eqref{Ropt}, with $h(\myvec{r})$ the mirror surface height],
we can re-express the eigenvalue problem as:
\begin{equation}
\mathcal{L}_{[h(\myvec{r})]}u \equiv \mathcal{G} \mathcal{R}_{[h(\myvec{r})]} \mathcal{G} u = \gamma u\,,
\end{equation}
with $\mathcal{L}_{[h(\myvec{r})]}$ the center-to-center
propagator when the mirror deviates from flat surfaces by
$h(\myvec{r})$, in which the optical mode propagates from the
cavity center to the mirror, gets reflected, and propagates back
to the center. In fact,  $\mathcal{L}$ is related to the
surface-to-surface propagator $\mathcal{K}$ by a unitary
transformation,
\begin{equation}
\label{commute:2}
\mathcal{L} = \mathcal{G}^{-1}\mathcal{R}_{\left[h(\myvec{r})/2\right]}^{-1} \mathcal{K}\mathcal{R}_{\left[h(\myvec{r})/2\right]}\mathcal{G}\,.
\end{equation}
This means the two proofs are mathematically equivalent. Similar
to $\mathcal{K}$, the operator $\mathcal{L}$ also commutes with
parity, or [Cf.~Eq.~\eqref{commute}]
\begin{equation}
\label{eq:parity:commute}
\mathcal{P}\mathcal{L}=\mathcal{L}\mathcal{P}\,;
\end{equation}

With the propagator on hand, we proceed with our intuition that
the modes must be related by Fourier transforms.  In order to do
so, we first define the 2-D Fourier-transform operator
$\mathcal{F}$ as
\begin{equation}
\mathcal{F}(\myvec{r},\myvec{r}')=\frac{k}{\pi L} e^{-\frac{2i k}{L}\myvec{r}\cdot\myvec{r}'}\,,
\end{equation}
which satisfies
\begin{equation}
\mathcal{F}^2=(\mathcal{F}^{-1})^2=\mathcal{P}\,.
\end{equation}
It is easy to show that,
\begin{eqnarray}
&&\left[\mathcal{G}^* \mathcal{F}^{-1}\right](\myvec{r},\myvec{r}\,') \nonumber \\
&=&
- \frac{i k^2}{\pi^2 L^2 }e^{ i k
\left[\frac{L}{2}+\frac{\myvec{r}^2}{L}+
\frac{(\myvec{r}-\myvec{r}')^2}{L}\right]}
\int d^2\myvec{r}\,'' e^{\frac{i k}{L}[\myvec{r}\,''-(\myvec{r}-\myvec{r}\,')]^2}\nonumber \\
&=&
\left[i e^{i k L}\mathcal{R}_{[\myvec{r}^2/(2L)]}\mathcal{G}\right](\myvec{r},\myvec{r}\,')\,.
\label{GF}
\end{eqnarray}
[The integral on the second line can be done by inserting a factor
$e^{-\epsilon (\myvec{r}\,'')^2}$ into the integrand, and then
letting $\epsilon \rightarrow 0^+$.] Similarly, [or by taking the
transpose of Eq.~\eqref{GF}], we have
\begin{equation}
\label{FG}
\mathcal{F}^{-1}\mathcal{G}^*i e^{i k L}\mathcal{G R}_{[\myvec{r}^2/(2L)]}\,.
\end{equation}

Using Eqs.~\eqref{GF} and \eqref{FG}, we have
\begin{eqnarray}
\label{identity}
&&\mathcal{P}\mathcal{L}_{[h_{\mathrm{A}}]}^* \nonumber \\
&=& \mathcal{F}^{-1} (\mathcal{F}^{-1} \mathcal{G}^*)
\mathcal{R}_{[-h_{\mathrm{A}}]}(\mathcal{G}^* \mathcal{F}^{-1})\mathcal{F} \nonumber\\
&=& - e^{ 2 i k L}\mathcal{F}^{-1} \mathcal{G R}_{[\myvec{r}^2/(2L)]}
\mathcal{R}_{[-h_{\mathrm{A}}]}\mathcal{R}_{[\myvec{r}^2/(2L)]}\mathcal{ G} \mathcal{F}\nonumber\\
&=&- e^{2 i k L}\mathcal{F}^{-1}  \mathcal{L}_{[h_{\mathrm{B}}]}\mathcal{F}\,.
\end{eqnarray}
Here $h_{\rm A}$ and $h_{\rm B}$ are mirror heights related by the duality relation,
\begin{equation}
\label{eq:mirror:duality}
h_A(\myvec{r})+h_B(\myvec{r}) = r^2/L\,,
\end{equation}
and we have  used the fact that
\begin{equation}
\mathcal{R}_{[\myvec{r}^2/(2L)]}
\mathcal{R}_{[-h_{\mathrm{A}}]}\mathcal{R}_{[\myvec{r}^2/(2L)]}
=\mathcal{R}_{[\myvec{r}^2/L-h_{\mathrm{A}}]}
=\mathcal{R}_{[h_{\mathrm{B}}]}\,.
\end{equation}

According to Eq.~\eqref{identity}, given any eigenstate $u_A$ of
$\mathcal{L}_{[h_{\mathrm{A}}]}$ with eigenvalue $\gamma_{\rm A}$
and a definite parity of $p$, we have
\begin{eqnarray}
(-1)^p \gamma_{\rm A}^* u_A^* &=& \mathcal{P}\mathcal{L}_{[h_{\mathrm{A}}]}^* u_A^* \nonumber \\
&=&
- e^{2 i k L} \mathcal{F}^{-1}\mathcal{L}_{[h_{\mathrm{B}}]} (\mathcal{F}u_A^*)\,,\\
\Rightarrow
\mathcal{L}_{[h_{\mathrm{B}}]} (\mathcal{F}u_A^*)& =& (-1)^{p+1}  e^{-2 i k L} \gamma_{\rm A}^*  (\mathcal{F}u_A^*)\,.
\end{eqnarray}
In other words, the mapping
\begin{eqnarray}
\label{mapping}
\label{fourier:mapping}
u_{\rm A}  &\rightarrow& u_{\rm B} = \mathcal{F}u_A^*
\end{eqnarray}
transforms each eigenstate of  $\mathcal{L}_{[h_{\mathrm{A}}]}$
into its dual one of  $\mathcal{L}_{[h_{\mathrm{B}}]}$; the
corresponding eigenvalue relation is
\begin{equation}
\gamma_{\rm B}= (-1)^{p+1}  e^{-2 i k L} \gamma_{\rm A}^*\,.
\end{equation}

For similar reasons, given any eigenstate $u_{\rm B}$ of
$U_{[h_{\mathrm{B}}]}$ (with definite parity), $\mathcal{F} u_{\rm
B}^*$ must also be an eigenstate of  $U_{[h_{\mathrm{A}}]}$.
Moreover, since
\begin{equation}
\mathcal{F}(\mathcal{F}u_{\rm B}^*)^*=\mathcal{F} \mathcal{F}^{-1}u_{\rm B}=u_{\rm B}\,,
\end{equation}
the state $\mathcal{F}u_{\rm B}^*$ is in fact the inverse image of
$u_{\rm B}$ [under the mapping \eqref{mapping}]. This means we
have established a one-to-one correspondence between eigenstates
of $\mathcal{L}_{[h_{\mathrm{A}}]}$ and those of
$\mathcal{L}_{[h_{\mathrm{B}}]}$.

Now let us look at intensity profiles on the end mirrors surface.
For the eigenstate $u_{\rm A}$, the field amplitude at the
constant-$z$ plane of the end mirror is $\mathcal{G}u_{\rm A}$.
For its image eigenstate $u_B\equiv \mathcal{F}u_{\rm A}^*$, we
have
\begin{eqnarray}
\mathcal{G} u_{\rm B} = \mathcal{G} (\mathcal{F}u_{\rm A}^*) &=& \left[\mathcal{G}^* \mathcal{F}^{-1}u_{\rm A} \right]^*\nonumber \\
&=&\left[i e^{i k L} R_{[\myvec{r}^2/(2L)]} \mathcal{G} u_{\rm A}\right]^*
\end{eqnarray}
which does have the same intensity profile [see Eq.~\eqref{Ropt}].

For the round-trip propagator $\mathcal{L}^2$, using Eqs.~\eqref{identity} and \eqref{eq:parity:commute}, we have
\begin{equation}
\left[\mathcal{L}_{[h_{\mathrm{A}}]}^2\right]^* =e^{4ikL}\mathcal{F}^{-1}\mathcal{L}_{[h_{\mathrm{B}}]}^2\mathcal{F}\,,
\end{equation}
so we have the same duality correspondence \eqref{mapping} between eigenstates of the full propagator, with the mapping between eigenvalues given by
\begin{equation}
\eta_{\rm B}=   e^{- 4 i k L} \eta_{\rm A}^*\,.
\end{equation}

\section{\label{sec:examples} Application of the duality relation using Mesa Beams and Mexican-Hat cavities}

The Mesa beams were constructed to have flat-topped intensity
profiles at the cavity mirrors with rapid fall-off near mirror
rims, in order to achieve lower thermal noises~\cite{osv,mbi,tnA}.
There are two versions of Mesa beams with the same intensity
profile, the nearly flat and the nearly concentric. Cavities that
support them (Mexican-Hat cavities) are related by the duality
relation, as realized by Savov~\cite{SandV}, during his study of
radiation-pressure-induced tilt instabilities. In this section, we
shall explicitly construct these two fundamental modes, study
their relations at the center of the cavity, and at the cavity
mirrors. We will also discuss analytical features of the two modes
that have not been obtained before.  We will also give an example
of how the calculation of the tilt instability can be dramatically
simplified for nearly concentric Mexican-hat cavities, using the
duality relation, based on results already obtained for the nearly
flat configuration.

\subsection{\label{subsec:examples:mapping} Construction of Mesa beams in Cartesian coordinate system}

Nearly-flat Mesa beams are constructed by coherently superimposing
minimal Gaussians, namely Gaussian modes with the smallest
possible spot size at the cavity mirrors,  $\sigma_{\rm
min}=\sqrt{L/(2k)}$, whose axes are parallel to the cavity axis
and lie within a cylinder centered at the cavity axis. At the
middle of the cavity, the axes intercept with the constant-$z$
plane in a disk $\mathcal{D}$, with radius $p$. It is evident that
such a construction will give a rather flat intensity profile in
the central region of the end mirror with radius $\sim p$; beyond
this radius, the intensity profile falls off as a Gaussian with
decay length $\sigma_{\rm min}$~\cite{osv,mbi}.

The complex amplitude of the nearly-flat Mesa beam (fundamental
mode of the corresponding cavity) at the center of the cavity is
of the form
\begin{equation}
\label{mesaA} v_{\rm f}(\myvec{r})= \int_{\myvec{r}_0 \in
\mathcal{D}} d^2 \myvec{r}_0 \,
\left(\frac{1}{\sqrt{2\pi}\sigma}\right)^2e^{-\frac{(\myvec{r}-\myvec{r}_0)^2}{2\sigma^2}}\,,
\end{equation}
Here $\sigma$ is the waist size, which we leave general (rather than setting $\sigma=\sigma_{\rm min}$) for the moment. The duality image of $v_{\rm f}$ is
\begin{eqnarray}
v_{\rm c}(\myvec{r})&=&\left[\mathcal{F}v_{\rm f}^*\right](\myvec{r})\nonumber \\
&=&\int_{\myvec{r}_0 \in \mathcal{D}}
d^2 \myvec{r}_0\,
e^{\frac{2ik\myvec{r}\cdot\myvec{r}_0}{L}}
\mathcal{F}\left[\left(\frac{1}{\sqrt{2\pi}\sigma}\right)^2e^{-\frac{\myvec{r}^2}{2\sigma^2}}\right]\nonumber \\
&=&\int_{\myvec{r}_0 \in \mathcal{D}}
d^2 \myvec{r}_0\,
e^{\frac{2ik\myvec{r}\cdot\myvec{r}_0}{L}}
\left[\left(\frac{1}{\sqrt{2\pi}\sigma_*}\right)^2
e^{-\frac{\myvec{r}^2}{2\sigma_*^2}}\right]\,,\quad\;
\label{mesaB}
\end{eqnarray}
with
\begin{equation}
\label{sigma}
\sigma \sigma_* = \frac{L}{2k}=\sigma_{\rm min}^2\,,
\end{equation}
When going from Eq.~\eqref{mesaA} to Eq.~\eqref{mesaB}, the
Fourier transform has been completed by two steps. First, the
spatial translation by $\myvec{r}_0$ is replaced by the phase
factor of $e^{\frac{2ik\myvec{r}\cdot\myvec{r}_0}{L}}$, which
represents a tilt of the propagation axis by an angle of
$2\myvec{r}_0/L$. Second, the $\sigma$-Gaussians turn into
$\sigma_*$-Gaussians. [This correspondence between Gaussians in
fact reflects the duality between pairs of spherical cavities.] As
a consequence, $v_{\rm c}$ represents the superposition of
Gaussians with symmetry axes going through the cavity center, but
with tilt angles distributed uniformly in a disk with radius
$2p/L$ --- exactly the construction of a nearly-concentric Mesa
beam. In particular, Eq.~\eqref{sigma} tells us that minimal
Gaussian would have turned into itself in this process. Hence we
have shown explicitly the correspondence between the nearly-flat
and nearly-concentric Mesa beams (the fundamental modes of the
corresponding cavities).

\subsection{\label{subsec:examples:profile} Profiles of Mesa beams and mirror shapes}

\begin{figure}[ht]
\begin{center}
\includegraphics[width=10cm]{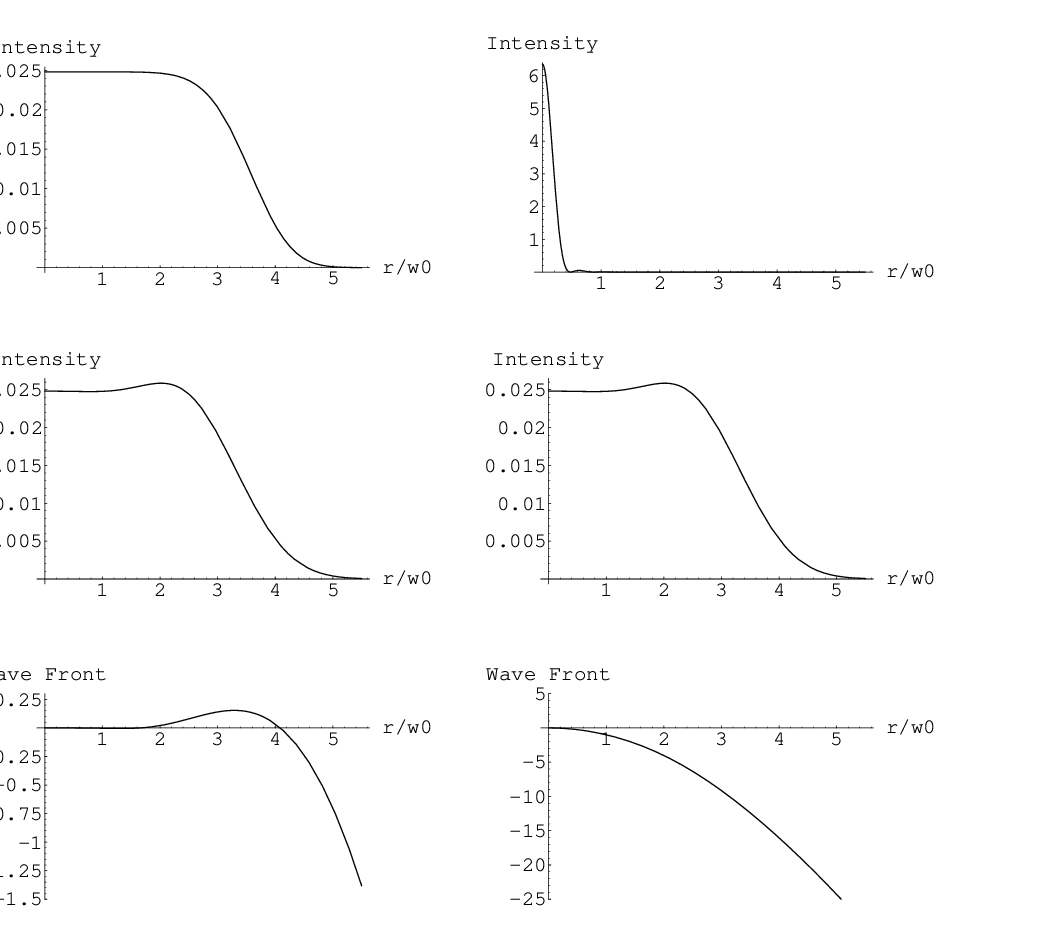}
\end{center}
\caption{Comparison between nearly flat (left panels) and nearly
concentric (right panels) Mesa beams. Upper panels: normalized
intensity profiles at the center of the cavity. Middle panels:
normalized intensity profiles at mirror surfaces Lower panels:
phase fronts at the position of the mirrors. \label{fig:aa}}
\end{figure}

In order to study Mesa beams in more details, we adopt the
cylindrical polar coordinate system $(r,\phi)$; the cylindrical
symmetry of these beams will make the complex amplitude only
depend on $r$. Equations~\eqref{mesaA} and \eqref{mesaB}, written
in the polar coordinate system, become
\begin{eqnarray}
v^{\rm waist}_{\rm f}(r,\phi)\!\!&=&\!\!
\frac{1}{\pi w_0^2}\int_{0}^{p}r_0\mbox{d}r_0 \nonumber \\
&& \!\!\int_{0}^{2\pi}\mbox{d}
\phi_0 \,  e^{-\frac{\scriptstyle r^2-2r_0r\cos(\phi-\phi_0)+r_{0}^{2}}{\scriptstyle w_{0}^{2}}},\quad\;\\
v^{\rm waist}_{c}(r,\phi)\!\!&=&\!\!
\frac{1}{\pi w_0^2}\int_{0}^{p}r_0
\mbox{d}r_0 \nonumber \\
&&\!\! \int_{0}^{2\pi}\mbox{d}\phi_0 \,
 e^{-\frac{\scriptstyle r^2+2ir_0r\cos(\phi-\phi_0)}
{\scriptstyle w_{0}^{2}}}\,.
\end{eqnarray}
Here $w_0=\sqrt{L/k}=\sqrt{2}\sigma_{\rm min}$ and $L$ is the
total length of the cavity. Carrying out the angular integrations
analytically, we get
\begin{eqnarray}
\label{flat_waist_radial}
v^{\rm waist}_{\rm f}(r) & = & \int_0^{p/w_0} 2x_0 e^{-(x^2+x_0^2)} I_0(2 x x_0) dx_0\,,\quad \\
\label{conc_waist_radial}
v^{\rm waist}_{\rm c}(r) & = & \frac{1}{x}e^{-x^2} J_1(2 x p/w_0)\,,
\end{eqnarray}
where $x\equiv r/w_0$, and $I_0$ is the modified Bessel function
of the first kind. Examples of normalized power distributions of
nearly flat and nearly concentric Mesa beams are plotted in the
upper panels of Fig.~\ref{fig:aa}. In these plots, we take
$p=4w_0$, which corresponds to the configuration proposed for
Advanced LIGO (for reasons that will be explained in
Sec.~\ref{subsec:examples:tilt}).

Let us analyze these amplitude distributions in more details, in
the case of $p \gg w_0$, i.e., when we translate the minimal
Gaussians by a distance substantially greater than their waist
size.  For the nearly-flat configuration, we can easily see from
Eq.~\eqref{mesaA} that, when $(p-r)/w_0 \gg 1$, the field
distribution can be approximated as
\begin{equation}
v_{\rm f}(r \ll p) \approx  \int_{\myvec{r}_0 \in \mathbf{R}^2}
d^2 \myvec{r}_0 \,
\left(\frac{1}{\sqrt{2\pi}\sigma}\right)^2e^{-\frac{(\myvec{r}-\myvec{r}_0)^2}{2\sigma^2}}=1\,.
\end{equation}
On the other hand, if $r$ is much larger than $w_0$ [since $p\gg
w_0$, this region overlaps with the previous one], we can apply
the asymptotic expansion of $I_0$
\begin{equation}
I_0(z)=\frac{1}{\sqrt{2 \pi z}}e^{z}
\end{equation}
on Eq.~\eqref{flat_waist_radial}, and obtain
\begin{eqnarray}
\label{vfasym}
&&v^{\rm waist}_{\rm f}(r \gg w_0) \nonumber \\
&  \approx &   \frac{1}{\sqrt{\pi}} \int_0^{p/w_0} \sqrt{\frac{x_0}{x}}e^{-(x_0-x)^2}dx_0\nonumber\\
&\approx & \frac{1}{\sqrt{\pi}}\int_{-x}^{p/w_0-x}\left(1+\frac{y}{2x}\right)e^{-y^2}dy\nonumber\\
&\approx&\left[\frac{1}{\sqrt{\pi}}\int_{-\infty}^{p/w_0-x}e^{-y^2}dy\right]-\frac{1}{4\sqrt{\pi}x}e^{-(p/w_0-x)^2}\,,\qquad
\end{eqnarray}
where we have defined $x\equiv r/w_0$. From Eq.~\eqref{vfasym}, we
note that when $w_0 \ll r \ll p$, we recover the result of
$v_f^{\rm waist} \approx 1$; when $r$ gets close to $p$, the
amplitude will drop, similar to the tail of an error function.
Qualitatively, we could write $w_{\mbox{\scriptsize f-Mesa}}(p)
\sim p$.  In the ultimate limit of $p/w_0\rightarrow +\infty$, we
have
\begin{equation}
v_{\rm f}^{\rm waist}(r)=1\,,\quad {p/w_0\rightarrow +\infty}\,.
\end{equation}

The concentric configuration, on the other hand, has a completely
different field distribution. According to the analytic
expression~\eqref{conc_waist_radial}, the amplitude must be
distributed within a radius of $x \sim w_0/p \ll 1$, or $r\sim
w_0^2/p$, which is much smaller than the waist size of the minimal
Gaussian. In this case, we could also qualitatively write
$w_{\mbox{\scriptsize c-Mesa}}(p) \sim w_0^2/p$. In the limit of
$p\rightarrow \infty$, we use
\begin{equation}
\frac{J_1(ax)}{x}\rightarrow \delta(x)\,,\quad a\rightarrow +\infty
\end{equation}
and have
\begin{equation}
v_{\rm c}^{\rm waist}(r)=\delta(x)\,,\quad {p/w_0\rightarrow +\infty}\,.
\end{equation}
The fact that
\begin{equation}
w_{\mbox{\scriptsize f-Mesa}}(p) \cdot w_{\mbox{\scriptsize c-Mesa}}(p) \sim w_0^2 \,,
\end{equation}
clearly reflects the Fourier-transform relation between two Mesa beams with the same $p$.

Now, let us turn to field distributions at  the cavity mirrors.
Applying the propagator between parallel planes in the polar
coordinate systems (eq.~\eqref{defG}),
\begin{eqnarray}
&& G(r',\phi';r,\phi)\nonumber \\
\!\!&=&\!\!\frac{ik}{\pi L}e^{-ikL/2}
e^{-{ik}[r^2+r'^2-2rr'
\cos(\phi'-\phi)]/L},\qquad
\end{eqnarray}
we obtain the fields
\begin{eqnarray}
\label{vendf} v^{\rm end}_{\rm
f}(r',\phi')\!\!&=&\!\!\int_{0}^{p}r_0\mbox{d}r_0
\int_{0}^{2\pi}\mbox{d}\phi_0 \nonumber \\ && e^{
-\left[\frac{\scriptstyle  1+i}{\scriptstyle 2}\right]\left[\frac{
\scriptstyle r'^2-2r_0r'\cos(\phi'
-\phi_0)+r_{0}^{2}}{\scriptstyle  w_{0}^{2}}\right]}\,, \\
\label{vendc}
v^{\rm end}_{\rm c}(r',\phi')\!\!&=&\!\!\int_{0}^{p}r_0\mbox{d}r_0  \int_{0}^{2\pi}\mbox{d}\phi_0 \nonumber \\
&& e^{-\left[\frac{\scriptstyle  1+i}{\scriptstyle  2}\right]\left[\frac{\scriptstyle  r'^2+2
ir_0r'\cos(\phi'-\phi_0)-ir_{0}^{2}}{\scriptstyle  w_{0}^{2}}\right]}\,,\quad\quad
\end{eqnarray}
at distance $L/2$ from the waist. Comparing Eqs.~\eqref{vendf} and \eqref{vendc}, we have
\begin{equation}
\label{vfvc}
\left[v_{\rm f}^{\rm end}(\myvec{r})\right]^* = e^{ik\myvec{r}^2/L}v_{\rm c}^{\rm end}(\myvec{r})\,.
\end{equation}
It is then obvious that the two beams have the same intensity profiles on the cavity mirrors:
\begin{equation}
\label{vfavca}
 |v_{\rm f}^{\rm end}(\myvec{r})|= |v_{\rm c}^{\rm end}(\myvec{r})|\,.
\end{equation}
(An approximate formula for the end-mirror intensity profile was
given in the Appendix of~\cite{osv}.) We plot these intensity
profiles at the mirror surfaces in the middle panels of
Fig.~\ref{fig:aa}.

\begin{figure}[ht]
\includegraphics[width=10cm]{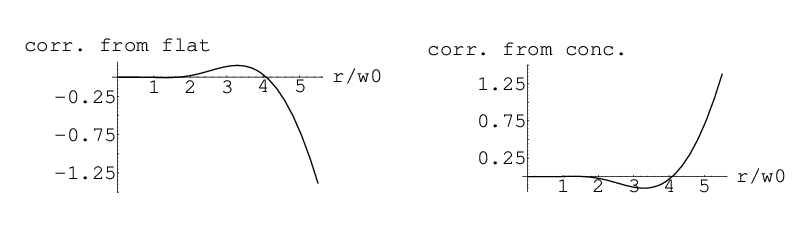}
\caption{Flat Mesa beam wave front (left panel) with respect to a
flat surface and concentric Mesa beam wave front (right panel)
with respect to a concentric surface, as analytically computed.
\label{fig:ab}}
\end{figure}

Let us now determine mirror shapes by imposing that the optical
phase is constant (which we take as 0 for simplicity) on each
mirror surface. We have
\begin{eqnarray}
\label{vfid}
v_{\rm f}^{\rm end}(\myvec{r})e^{ikh_{\rm f}(\myvec{r})} &=& |v_{\rm f}^{\rm end}(\myvec{r})|\,, \\
\label{vcid}
v_{\rm c}^{\rm end}(\myvec{r})e^{ikh_{\rm c}(\myvec{r})} &=&
|v_{\rm c}^{\rm end}(\myvec{r})|
\,.\qquad
\end{eqnarray}
Taking the complex conjugate of Eq.~\eqref{vfid}, and combine with Eq.~\eqref{vcid}, using Eqs.~\eqref{vfvc} and \eqref{vfavca}, we have
\begin{equation}
h_{\rm f}(\myvec{r}) = \frac{\myvec{r}^2}{L} - h_{\rm c}(\myvec{r})\,,
\end{equation}
which is the duality relation between mirror surfaces. In the
lower panels of Fig.~\ref{fig:ab}, we plot the shapes of mirror
surfaces, again, we assume $p=4w_0$.

\subsection{Applications of Mesa beams to Advanced LIGO}
\label{subsec:examples:tilt}

In order to achieve lower thermal noise in the test masses, the
intensity profiles at the mirrors must be as flat as possible. In
the case of infinite mirrors, the choice is to use cavities with
flat or concentric spherical mirrors, whose eigenmodes have
uniform (absolutely flat) profile distribution. However, the
mirrors must have finite sizes (e.g., as limited by the size of
the beam tube), and the intensity profiles must be confined to a
very large extent within the rims of the mirrors, in order to
decrease the diffraction loss upon each reflection.  In Advanced
LIGO, a power loss below 10\,ppm is required~\cite{osv}. For this
reason, we are forced to deviate from flat or concentric
configurations --- to such an extent that the diffraction loss is
within the requirement. When only spherical mirrors are used, if
on the one hand we decrease the radius of curvature from $+\infty$
(flat), and on the other hand increase the radius of curvature
from $L/2$ (concentric), the dual configurations, with
\begin{equation}
1/(2R_1)+1/(2R_2)=1/L\,,
\end{equation}
will have the same intensity profiles at the end mirrors, thus the
same diffraction loss and thermal noise. For example, $R_1=54\,$km
and $R_2=2.077\,$km both give exactly the loss specification,
while $R_1$ is the current baseline design. However, spherical
cavities are not optimal in terms of their thermal noise: (the two
types of) Mesa beams, whose intensity profiles are flatter given
the same loss specification, turn out to provide much lower
thermal noises~\cite{osv,tnA}. For these beams, the larger the
parameter $p$, the lower the thermal noises, but the higher the
diffraction loss. The loss specification of Advanced LIGO
corresponds to $p=4w_0$ \cite{osv} which is the case we study in
Fig.~\ref{fig:aa}.

While having the same diffraction losses and thermal noises, dual
configurations do differ significantly in a very important aspect
--- their eigenspectra are different. Thus, any problem using
modal analysis of optical cavities will reveal these differences
and probably the duality relation if nearly flat and nearly
concentric configurations are compared.

\begin{figure}[ht]
\begin{center}
\includegraphics[width=8cm]{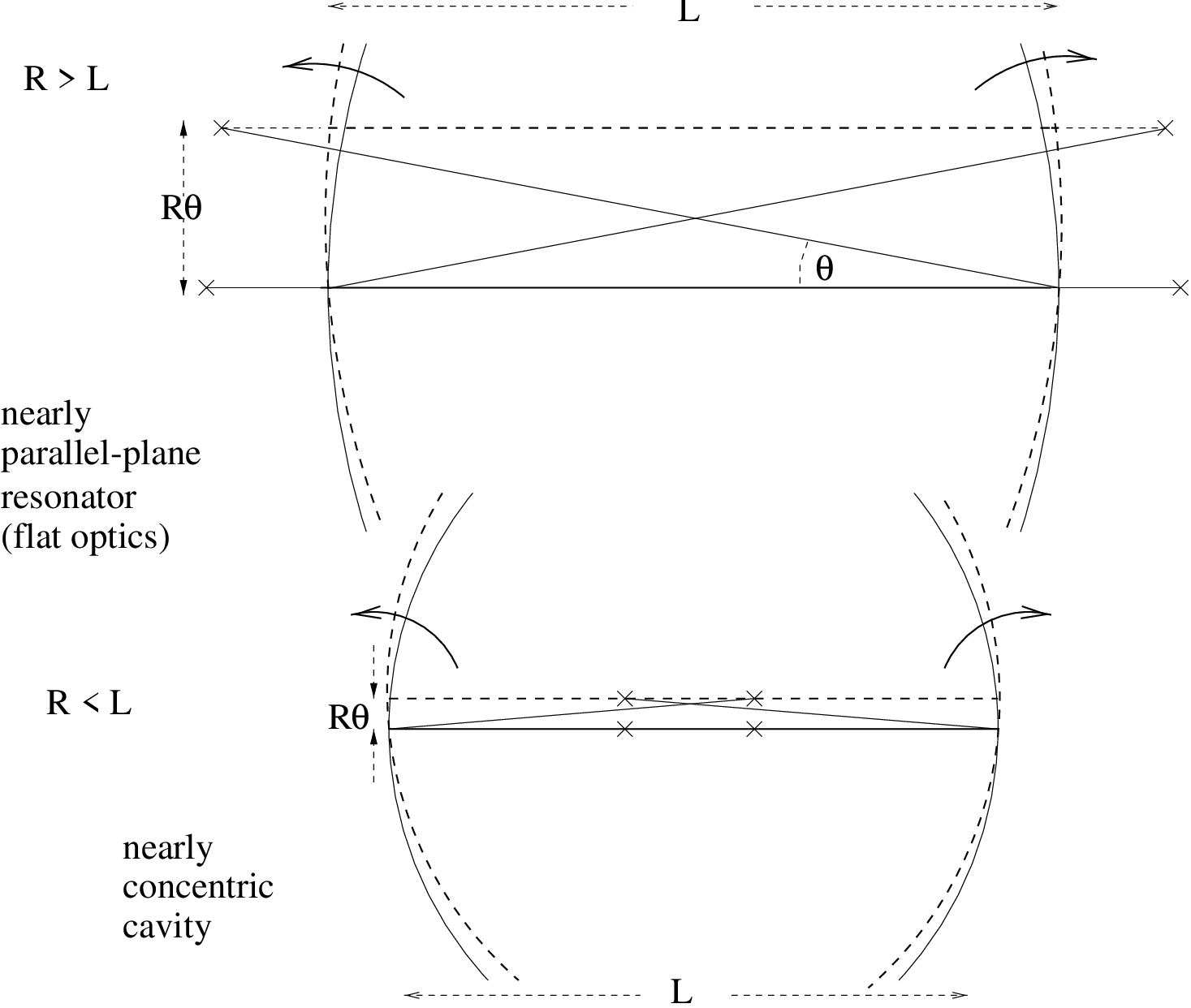}
\end{center}
\caption{Comparison of tilt instability of nearly flat and nearly
concentric symmetric optical cavities. For more details see
Ref.~\cite{sigg,SandV}. \label{fig:ac}}
\end{figure}

One such problem is the radiation-pressure-induced tilt
instability: as the mirrors tilt, the beam inside the cavity walks
away from the center of the mirrors, producing a torque, which in
some cases can drive more tilt in the same direction, and become
destabilizing (see Fig.~\ref{fig:ac}). As shown by
Sigg~\cite{sigg}, while for all cavities there is always one tilt
mode in which the radiation-pressure-induced torque is
destabilizing, the instability is much weaker in nearly concentric
configurations than in nearly flat ones. The reason is that while
in the two cases the intensity profiles are identical, the optical
axis of the beam walks away by a much smaller distance in the
concentric case, given the same amount of tilt in the unstable
mode (see Fig.~\ref{fig:ac}). According to Sigg's calculation for
spherical mirrors, the tilt instability for a nearly flat
configuration with Advanced-LIGO power ($\sim 1\,$MW circulating
in the cavity) can be too strong to handle for the angular control
system. For this reason, we would prefer nearly concentric
cavities.

For general, non-spherical cavities, a perturbative prescription
for calculating the tilt instability has been formulated by Savov
and Vyatchanin~\cite{SandV}, in which the tilt instability growth
time is expressed in terms of eigenvalues and intensity profiles
of the cavities' spatial eigenmodes (Eqs.~2.13, 2.14, and 4.8
of~\cite{SandV}). Savov and Vyatchanin applied their prescription
to both nearly flat and nearly concentric Mexican-Hat cavities; in
particular, they had to solve the eigenvalue problem for the
nearly concentric cavities in order to obtain the eigenvalues and
intensity profiles. Savov discovered the duality relation in this
process. Had the duality relation been known, one could have taken
the eigenvalues and intensity profiles of nearly flat Mexican-hat
cavities, available from previous works, applied the duality
transformation, and obtained the tilt instability for nearly
concentric Mexican-Hat cavities without having to solve the
eigenvalue problem again (see Section VI of~\cite{SandV}).

Finally, let us make a qualitative comment on the numerical
magnitudes of tilt instabilities in the various configurations
considered. Numerically, according to Savov and
Vyatchanin~\cite{SandV}, we have
\begin{equation*}
\begin{array}{c}
\mbox{nearly flat MH ($p=4w_0$)} \\
\downarrow \\
\mbox{nearly flat spherical ($R=54\,$km)} \\
\downarrow \\
\mbox{nearly concentric spherical ($R=2.077\,$km)}\\
\downarrow \\
\mbox{nearly concentric MH ($p=4w_0$)}\\
\end{array}
\end{equation*}
with configurations less and less unstable from top to bottom.
Interestingly, this sequence of decreasing instability is
consistent qualitatively with the corresponding mirror shapes:
with the same amount of diffraction loss, the flat MH does appear
more flat than the nearly flat spherical mirrors, while the nearly
concentric Mexican-Hat mirror does appear closer to concentric
than the nearly concentric spherical mirror.

\section{Conclusion}
\label{sec:conclusion} In this paper, we provided two different
analytic proofs for a duality relation between symmetric cavities
with mirror height $h(\myvec{r})$ measured with respect to a flat
surface and those with mirror height $-h(\myvec{r})$ measured with
respect to a concentric spherical surface ({\it valid within the
paraxial approximation}): the corresponding eigenmodes have the
same intensity profile at the mirrors, their amplitude
distribution at the center of the cavity is related via Fourier
transform, while their eigenvalues are related by complex
conjugation (see Table~\ref{tab:duality:EJ}). These two proofs are
based on the mirror-to-mirror propagator, and the center-to-center
propagator, respectively.

We illustrated this duality relation with the two types of Mesa
beams proposed for Advanced LIGO. In particular, we showed
explicitly that these beams are related to each other by a Fourier
transform at the center of the cavities, and that they have the
same intensity profiles at the end of the cavities. We also
related the mirror shapes of the Mexican-Hat cavities that support
these two modes by the duality relation. The duality relation
could have allowed us to avoid solving the eigenequations once
more for the nearly concentric Mexican-Hat cavities, and used
instead results already available for nearly flat Mexican-Hat
cavities.

In addition, this duality relation can also be applied to more
general optical cavities, which interpolate between nearly flat
and nearly concentric ones~\cite{BT}.

This duality relation provides a quite general tool for designing
non-standard optical cavities and for studying the performances of
dual configurations. Several physical effects such as coupling of
optical cavities with particles ~\cite{OM}, coupling of mechanical
and optical degrees of freedom in optical resonators ~\cite{PI},
cavity misalignment sensitivity, depend on the optical cavity
modes structure in terms of beam geometry and eigenvalues.
Therefore this unique mapping might be useful not only for
studying practical issues related to advanced ground-based
gravitational waves interferometers such as the parametric
instability ~\cite{PIMB}, but in a variety of other applications
which could benefit from using Fabry-Perot cavities with
non-spherical mirrors, including precision metrology and atomic
physics. Ultrastable optical cavities have become a standard tool
for stabilizing laser systems needed for example for
high-resolution spectroscopy and optical clocks. Current cavities
~\cite{USL} are mostly limited by the mirrors thermal noise and
this may be reduced, inter alia, by using non spherical mirrors
supporting non-Gaussian beam. Particles manipulation in optical
resonators is based on electric dipole interaction with the  laser
fields (optical dipole traps ~\cite{ODT}) and the  potential
energy of the induced dipole force is related to the intensity
distribution of the laser beam. Non spherical mirrors could be
employed for the optimization of the geometry and depth of the
optical potential.

\acknowledgements

We thank K.S.~Thorne and W.~Kells for useful discussions. Research
of P.S.\ and Y.C.\ was supported by National Science Foundation
under Grant No. PHY-0099568.  Research of Y.C.\ was also sponsored
by the Alexander von Humboldt Foundation's Sofja Kovalevskaja
Award (funded by the German Federal Ministry of Education and
Research) and the David and Barbara Groce fund at the San Diego
Foundation. Research of J.A.\ and E.D'A.\ was supported by
National Science Foundation under Grant No. PHY-0107417.

\appendix

\section{Duality relation for non-identical mirrors}

In this section we will study the duality relation in the case of
not identical mirror shapes, but still symmetric under a
$180^\circ$ rotation around the cavity axis. Now the field
distributions of eigenstates over the two mirror surfaces are not
identical and we have to study the eigenvalue problem associated
with the round-trip propagator. Nevertheless, we can still use the
propagators~\eqref{eq:aa} and~\eqref{eq:aaconc} to build a system
of integral equations relating field distributions $v_1(\vec r_1)$
and $v_2(\vec r_2)$ over the two mirror surfaces. [All through
this section, we use the subscripts $1$ and $2$ to refer to
quantities associated with mirrors $1$ and $2$, respectively.] If
the mirrors deviate from parallel planes by $h_{1,2}(\vec r)$, we
have: \bea \label{noneq1}
&& \gamma_1 v_{1}(\vec r_{1}) = \int_{S_2} \mbox{d}^2\vec r_2 \;{\cal K}_{12}(\vec r_1 , \vec r_2)\, v_{2}(\vec r_{2})\,, \\
\label{noneq2}
&&\gamma_2  v_{2}(\vec r_{2}) = \int_{S_1} \mbox{d}^2\vec r_1\; {\cal K}_{21}(\vec r_2 , \vec r_1)\, v_{1}(\vec r_{1})\,,
\eea
where $\gamma_{1,2}$ are the ``eigenvalues'' and
\bea
&&{\cal K}_{12}(\vec r_1 , \vec r_2)=\frac{i k e^{-ikL} }{2\pi L}
 e^{ikh_1(\vec r_1)-\frac{ik}{2L}|\vec r_1-\vec r_2|^2 +ikh_2(\vec r_2)},\qquad\\
&&{\cal K}_{21}(\vec r_2 , \vec r_1)=\frac{i k e^{-ikL} }{2\pi L}
 e^{ikh_2(\vec r_2)-\frac{ik}{2L}|\vec r_2-\vec r_1|^2+ikh_1(\vec r_1)},\qquad
\eea

are the propagators from mirror $2$ to mirror $1$, and from mirror
$1$ to mirror $2$, respectively. The equations ~\eqref{noneq1} and
\eqref{noneq2} give the field at each mirror in terms of the
reflected field at the other but they can be combined to form the
round-trip equation which states that the field at each mirror
must reproduce itself after one round-trip. In the following, we
will add a subscript $\rm f$ or $\rm c$ to make a distinction
between quantities related to the nearly-flat or nearly-concentric
case.

 \bea
&& \eta_f \, v_{1f}(\vec r_1) = \int_{S_1 '} \mbox{d}^2\myvec r_1' \;{\cal K}_{1f}^{h_1h_2}(\vec r_1 , \myvec r_1')\, v_{1f}(\myvec r_1') ,\qquad \\
&& \eta_f \, v_{2f}(\vec r_2) =\int_{S_2 '} \mbox{d}^2\myvec r_2' \;{\cal K}_{2f}^{h_2h_1}(\vec r_2 , \myvec r_2')\, v_{2f}(\myvec r_2'), \qquad
\eea
where the common eigenvalue $\eta_f$ is given by $\gamma_{1f}\gamma_{2f}$ and the round-trip propagators
\bea
\label{eq:gen:f}
&&{\cal K}_{1f}^{h_1h_2}(\vec r_1 , \myvec r_1')=  \int_{S_2} \mbox{d}^2\vec r_2 \;
 {\cal K}_{12f}(\vec r_1 , \vec r_2)\, {\cal K}_{21f}(\vec r_2 , \vec r_1)\nonumber \\
&&{\cal K}_{2f}^{h_2h_1}(\vec r_2 , \myvec r_2')= (1\leftrightarrow 2)\cdot {\cal K}_{1f}^{h_1h_2}(\vec r_1 , \myvec r_1')
\eea

In the nearly-concentric configuration, using kernels of the form
~\eqref{eq:aaconc} for the propagation from one mirror to the
other and combining them as done for the nearly-flat
configuration, we obtain the following nearly-concentric
round-trip equation for the field distribution over the mirror $1$
(similar formula for the mirror $2$ with the substitution
$1\leftrightarrow 2$).

\bea \label{eq:gen:c}
&& \eta_c \, v_{1c}(\vec r_1) = \int_{S_1 '} \mbox{d}^2\myvec r_1' \;{\cal K}_{1c}^{b_1b_2}(\vec r_1 , \myvec r_1')\, v_{1c}(\myvec r_1') \\
&&{\cal K}_{1c}^{b_1b_2}(\vec r_1 , \myvec r_1')= - \int_{S_2} \mbox{d}^2\vec r_2 \; e^{-2ikL}\Big(\frac{k}{2\pi L}\Big)^2 \cdot\\
&&\cdot \,e^{\frac{ik}{2L}|\vec r_1+\vec r_2|^2+\frac{ik}{2L}|\vec
r_2+\myvec r_1'|^2+ikb_1(\vec r_1)+ikb_1(\myvec r_1')+2ikb_2(\vec
r_2)} \nonumber \eea

where $b_{1,2}$ are the mirrors deviations from concentric
surfaces. Using the assumed symmetry properties of the mirrors,
the propagators for the nearly-flat and nearly-concentric cavity
fulfills this relation (the same is true for the mirror $2$ with
the substitution $1\leftrightarrow 2$)

\bea \label{eq:gen:dual}
{\cal K}_{1c}^{-h_1-h_2}(\vec r_1 , \myvec r_1')&=& e^{-4ikL}[{\cal K}_{1f}^{h_1h_2}(-\vec r_1 , -\myvec r_1')]^* \nonumber \\
&=&  e^{-4ikL}[{\cal K}_{1f}^{h_1h_2}(\vec r_1 , \myvec r_1')]^*
\eea

Equation~\eqref{eq:gen:dual}, together with Eqs.~\eqref{eq:gen:f}
and \eqref{eq:gen:c}, express a general duality relation for
cavities with non-identical mirrors: as long as the corresponding
mirrors of two cavities $A$ and $B$ satisfy \beq h_{\alpha A}(\vec
r)=\frac{\vec r\,^2}{L} - h_{\alpha B}(\vec r)\,, \quad
\alpha=1,2\,, \eeq the eigenstates and eigenvalues of the two
cavities will be related by: \bea v_{\alpha A} = v_{\alpha
B}^*\,,\quad \eta_A = e^{-4ikL} \eta_B^*\,, \quad \alpha=1,2\,.
\eea

\section{Eigenstates and eigenvalues for cavities with infinite mirrors}
\label{app:b}

When the mirrors are infinite, it is straightforward to check that two fundamental properties,
\begin{eqnarray}
\int\mbox{d}^2\vec r\,' {\cal K}(\vec r,\vec r\,')
{\cal K}^*(\vec r\,',\vec r\,'') &=&\delta(\vec r-\vec r\,'')\,,
\\ {\cal K}(\vec r,\vec r\,') &=& {\cal K}(\vec r\,',\vec r)\,,
\end{eqnarray}
are satisfied by both propagators ${\cal K}_{\rm f}^{h}$ and
${\cal K}_{\rm c}^{b}$; they can be re-written into

\bea \label{K:relations}
\mathcal{K}\mathcal{K}^{\dagger}=\mathbf{I}, \quad
\mathcal{K}=\mathcal{K}^T\,,
\eea

where $\mathbf{I}$ is identity operator, $\mathcal{K}^T$ the
conjugate of $\mathcal{K}$, and $\mathcal{K}^{\dagger}$ its
Hermitian conjugate. In simple terms, $\mathcal{K}$ is unitary and
symmetric.   It is well known that for unitary operators, all
eigenvalues have modulus 1, and that eigenvectors with different
eigenvalues are orthogonal to each other.

Now suppose we have an eigenvector $v$, with eigenvalue $\gamma$,
$\gamma \gamma^*=1$. By complex conjugating the eigenequation
$\mathcal{K} v =\gamma v$, we obtain \beq \mathcal{K}^* v^* =
\gamma^* v^* = \gamma^{-1} v^*\,; \eeq using
Eqs.~\eqref{K:relations}, we have
$\mathcal{K}^*=\mathcal{K}^\dagger = \mathcal{K}^{-1}$, and hence
\beq \mathcal{K}^{-1} v^* = \gamma^* v^* \Rightarrow \gamma v^*
=\mathcal{K} v^*\,. \eeq This means $v^*$ and $v$ are both
eigenvectors with eigenvalue $\gamma$. We can then replace $v$ and
$v^*$ by two real eigenvectors of the eigenvalue problem, $v+v^*$
and $(v-v^*)/i$.  This corresponds to the physical fact that the
optical phase of eigenstates must be constant on mirror surfaces.

\end{document}